\documentclass[aps,prl,twocolumn,showpacs,superscriptaddress]{revtex4}
\usepackage{psfig}
\usepackage{epsfig}
\usepackage{array}
\usepackage{amssymb}
\usepackage{amsmath}
\usepackage{color}

\newcommand{\avg}[1]{\left<{#1}\right>}
\newcommand{\wdh}[1]{\widehat{#1}}
\newcommand{\wdt}[1]{\widetilde{#1}}

\begin{document}

\title{Theoretical derivation of $1/f$ noise in quantum chaos}
\author{E. Faleiro}

\affiliation{Departamento de F\'{\i}sica Aplicada,
	 E. U. I. T. Industrial, Universidad Polit\'ecnica de Madrid,
	 E-28012 Madrid, Spain}

\author{J. M. G. G\'omez}
\affiliation{Departamento de F\'{\i}sica At\'omica, Molecular y Nuclear, 
         Universidad Complutense de Madrid,  
         E-28040 Madrid, Spain}

\author{R. A. Molina}
\affiliation{CEA/DSM, Service de Physique de l'Etat Condens\'e, Centre
d'Etudes de Saclay, 91191 Gif-sur-Yvette, France}

\author{L. Mu\~noz}
\affiliation{Departamento de F\'{\i}sica At\'omica, Molecular y Nuclear, 
         Universidad Complutense de Madrid,  
         E-28040 Madrid, Spain}

\author{A. Rela\~no}
\affiliation{Departamento de F\'{\i}sica At\'omica, Molecular y Nuclear, 
         Universidad Complutense de Madrid,  
         E-28040 Madrid, Spain}

\author{J. Retamosa}
\affiliation{Departamento de F\'{\i}sica At\'omica, Molecular y Nuclear, 
         Universidad Complutense de Madrid,  
         E-28040 Madrid, Spain}

\begin{abstract}
It was recently conjectured that $1/f$ noise is a fundamental
characteristic of spectral fluctuations in chaotic quantum
systems. This conjecture is based on the behavior of the power
spectrum of the excitation energy fluctuations, which is different for
chaotic and integrable systems. Using random matrix theory we derive
theoretical expressions that explain the power spectrum behavior at
all frequencies. These expressions reproduce to a good approximation
the power laws of type $1/f$ ($1/f^2$) characteristics of chaotic
(integrable) systems, observed in almost the whole frequency
domain. Although we use random matrix theory to derive these results,
they are also valid for semiclassical systems.
\end{abstract}

\pacs{PACS:21.60.Cs,23.40.Bw}

\maketitle

% 
% ******************* INTRODUCTION ************************
% ********************  GENERAL    ************************

The study of energy level fluctuations is a basic tool in
understanding quantum chaos. The pioneering work of Berry and Tabor
showed that the spectral fluctuations of a quantum system whose
classical analogue is fully integrable are well described by Poisson
statistics \cite{Berry:77}. On the other hand, Bohigas {\it et al.}
conjectured that the fluctuation properties of quantum systems which
in the classical limit are fully chaotic coincide with those of Random
Matrix Theory (RMT) \cite{Bohigas:84}. Recently, a different approach
to quantum chaos has been proposed \cite{Relano:02}. Considering the
sequence of energy levels as a discrete time series, it has been
conjectured that chaotic quantum systems are characterized by $1/f$
noise, whereas integrable quantum systems exhibit $1/f^2$ noise. This
conjecture is supported by numerical experiments which involve
classical Random Matrix Ensembles (RME) and atomic nuclei
\cite{Relano:02}.

In this Letter we present a theoretical derivation of $1/f$ and
$1/f^2$ noises in chaotic and regular systems, respectively.  We use
RMT to derive these results, but they are also valid for semiclassical
systems. We present the main steps of the derivation and compare the
theoretical results with numerical calculation for RME, an atomic
nucleus and a quantum billiard, finding excellent agreement.

For any quantum system the accumulated level density $N(E)$ can be
separated into a smooth part $\overline{N}(E)$ and a fluctuating part
$\wdt{N}(E)$. To remove the main trend defined by the former, the
energy levels $E_i$ are mapped onto new dimensionless levels
$\epsilon_i = \overline{N}(E_i)$. This transformation, called
unfolding, allows to compare the statistical properties of different
systems or different parts of the same spectrum.

The analogy between the energy spectrum and a discrete time series is
established in terms of the $\delta_q$ statistic, defined as
\begin{equation}
\delta_q  = \sum_{i=1}^q (s_i - \left<s \right>) 
=\epsilon_{q+1}-\epsilon_1-q,
\label{def_delta}
\end{equation} 
where $\epsilon_i$ is the $i$-th unfolded level and $s_i =
\epsilon_{i+1} - \epsilon_i$. Once the unfolding is performed, the
average nearest level spacing is $\left<s\right> = 1$. Note that
$\delta_q$ represents the deviation of the excitation energy of the
$(q+1)$-th unfolded level from its mean value. Moreover, it is closely
related to the level density fluctuations. Indeed, we can write
\begin{equation}
\delta_q = -\tilde{N}(E_{q+1}), 
\label{def_delta_2}
\end{equation}
if we appropriately shift the ground state energy; thus, it represents
the accumulated level density fluctuations at $E = E_{q+1}$. Its power
spectrum, defined as the square of the modulus of the Fourier
transform, shows neat power laws both for fully chaotic and integrable
systems, i.e.,
\begin{equation}
P_k^{\delta} \propto \dfrac{1}{k^\alpha},
\label{1/fa}
\end{equation}
but level correlations change the exponent from $\alpha=2$ for
uncorrelated spectra to $\alpha = 1$ for chaotic quantum systems
\cite{Relano:02}.

{\em Notation.--} We consider an interval of length $L\gg 1$
containing $N\simeq L$ unfolded energy levels. The fluctuating parts
of the level and accumulated level densities are denoted
$\wdt{\rho}(\epsilon)$ and $\wdt{n}(\epsilon)$, respectively. In
addition to the $\delta_q$ statistic, we also introduce another
discrete function $\wdt{n}_q = \wdt{n}(q)$, obtained by sampling the
continuous function for integer values of the energy. The Fourier
transforms and power spectra of these functions are defined in the
usual way \cite{dft}.  The differences between continuous and discrete
functions, and specially between their Fourier transforms, play a
relevant role in the following. The notation we use for all these
functions, Fourier transforms and power spectra is summarized in Table
\ref{notation}.

\begin{table}
\caption[]{Summary of the functions, Fourier transforms and power
spectra used in this letter}
\begin{tabular*}{8cm}[c]{l|c|cc} \hline\hline
%\begin{tabular}{m{2cm}|p{2cm}|p{2cm}p{2cm}} \hline\hline
                       &                    &                                  \\[-5pt]
Domain                 &$\mathbb{R}$        & \multicolumn{2}{c}{$\mathbb{Z}$} \\[5pt]\hline
                       &                    &                                  \\[-5pt]
Function               &\makebox[1.8cm]{$\wdt{n}(\epsilon)$} &\makebox[1.8cm]{$\wdt{n}_q $}
  & \makebox[1.8cm]{$\delta_q$}       \\[5pt]
Fourier transform      &$\wdh{n}(\tau)$     & $\wdh{n}_k $  & $\wdh{\delta}_k$ \\[5pt]
Power spectrum         &$P^n(\tau)$         & $P^n_k$       & $P^{\delta}_k$   \\[5pt] \hline
\end{tabular*}
\label{notation}
\end{table}

Spectral and ensemble averages will be denoted by $\avg{\cdot}$, or
$\avg{\cdot}_\beta$ to distinguish ensemble averages performed in
different RME. Here, $\beta$ stands for the repulsion parameter
characterizing the ensemble. In this work we shall consider two RME:
the Gaussian Orthogonal Ensemble (GOE) and the Gaussian Unitary
Ensemble (GUE) \cite{Mehta:91}. We have $\beta=1$ for GOE and $\beta=2$
for GUE.

{\em Spectral fluctuations.--} The main object of this Letter is to
obtain explicit expressions of the average value of $P^{\delta}_k$ for
chaotic and integrable systems.  Except for integrable systems, one of
the main features of quantum spectra is that successive level spacings
are not independent random variables, but correlated quantities. This
property makes exceedingly difficult to work directly with the
discrete $\delta_q$ sequence. To circumvent this difficulty we profit
from the relationship (\ref{def_delta_2}). The statistical properties
of $\wdt{n}(\epsilon)$ are usually measured in terms of the spectral
form factor, defined as
\begin{equation}
K(\tau) = \avg{\left| \int d\epsilon \wdt{\rho}(\epsilon)
e^{-i 2\pi\epsilon\tau}\right|^2},
\end{equation}
that is, as the power spectrum of the fluctuating part of the energy
level density. Instead of $K(\tau)$, we can use the power spectrum
$P^n(\tau)$ of $\wdt{n}(\epsilon)$ to analyze spectral fluctuations.
It can be shown that for very large $L$ values \cite{Leboeuf:02}
\begin{equation}
\avg{P^n(\tau)} = \dfrac{K(\tau)}{4\pi^2\tau^2}.
\label{n_PWS}
\end{equation}

\noindent
Then, we shall (i) sample $\wdt{n}(\epsilon)$ for integer
values of the energy and study the fluctuations of the new $\wdt{n}_q
$ function; (ii) map the independent variable of $\wdt{n}_q$ (an
energy) to the independent variable of $\delta_q$ (a dimensionless
index).

{\em Quantum Chaos.--} One of the most important features of fully
chaotic system is the universal behavior of their spectral
fluctuations. Therefore, we can use RMT to obtain the main properties
of $\avg{P^n(\tau)}$ in this kind of systems. Exact analytical
expressions are known \cite{Mehta:91} for $K(\tau)$ in all RME,
leading to
\begin{eqnarray}
\avg{P^n(\tau)}_{\beta=2} &= & \left\{ \begin{array}{ll}
\dfrac{1}{4 \pi^2 \tau},   & \mbox{$\tau \le 1$}, \\
\dfrac{1}{4 \pi^2 \tau^2}, & \mbox{$\tau \ge 1$},
\end{array} 
\label{GUE}
\right. \\
\avg{P^n(\tau)}_{\beta=1} &= & \left\{ \begin{array}{ll}
\dfrac{2-\log(1 + 2 \tau)}{4 \pi^2 \tau}, & \mbox{$\tau \le 1$}, \\
\dfrac{2- \tau \log \left( \dfrac{2 \tau + 1}{2 \tau - 1} \right)}{4 \pi^2 \tau^2}, & \mbox{$\tau \ge 1$}.
\end{array} 
\right. 
\label{GOE}
\end{eqnarray}
For small $\tau$ values, (\ref{GUE}) and (\ref{GOE}) become (see Ref. \cite{Robnik:03})

\begin{equation}
\avg{P^n(\tau)}_{\beta} = \dfrac{1}{2\beta\pi^2\tau},\;\; \tau \ll 1,
\label{pn_smallt}
\end{equation}
and for $\tau\ge 1$ we can approximate $\avg{P^n(\tau)}_{\beta}$ by

\begin{equation}
\avg{P^n(\tau)}_{\beta} \simeq \dfrac{1}{4\pi^2\tau^2},\;\; \tau\ge 1,
\label{pn_tge1}
\end{equation}
which is exactly equal for GUE and a good approximation for GOE.

On the other hand, using periodic orbit theory and semiclassical
mechanics it is possible to calculate $K(\tau)$ for chaotic
systems. The semiclassical expression essentially coincides with the
results of RMT for $\tau_{min} \ll \tau \ll \tau_H$, where
$\tau_{min}$ is the period of the shortest periodic orbit, and
$\tau_H=h/\left<s\right>$ is the Heisenberg time of the system,
related to the time a wave packet takes to explore the complete phase
space of the system \cite{Ozorio:84,Berry:85}. The expressions derived
in the following are then directly applicable to generic chaotic
quantum systems for $\tau$ between this two values.

The next step is to relate the spectral fluctuations of $\wdt{n}_q$ to
those of the continuous function $\wdt{n}(\epsilon)$, as given by their
power spectra. With the usual definitions for the continuous and
discrete Fourier transforms we have
\begin{equation}
\wdh{n}_k = \dfrac{1}{\sqrt{N}}\sum_{q=-\infty}^{\infty}
\wdh{n}\left(\dfrac{k}{N}+q\right).
\label{aliasing}
\end{equation}
Therefore, to relate $P^n_k$ and $P^n(\tau)$ we need the precise
knowledge of $\wdh{n}(\tau)$ for all $\tau$. Since RMT only provides
the mean value of its squared modulus, we introduce two simplifying
assumptions: (a) for times $\tau=k/N$, $\left|\wdh{n}(k/N)\right| =
\sqrt{N \avg{P^n(k/N)}}$; (b) the phases of $\wdh{n}(k/N)$ are random
variables uniformly distributed in the interval $[0,2 \pi)$. These
assumptions, specially (a), are reasonable as far as we are only
interested in the average values of $P^n(\tau)$ and $P^n_k$.  With
these simplifications and using eq. (\ref{pn_tge1}) we obtain the
following result for $N\gg 1$,
\begin{equation}
\begin{split}
\avg{P^n_k} & = \avg{P^n(k/N)} +\avg{P^n(1-k/N)}-\dfrac{N^2}{4\pi^2k^2} \\
            & -\dfrac{N^2}{4\pi^2 (N-k)^2}+\dfrac{1}{4\sin^2\left(\dfrac{\pi k}{N}\right)},
              \;\; k=1,2,\cdots,N-1.
\end{split}
\label{pnk_pn}
\end{equation}

Finally, from the relationship between the variances of $\delta_q$ and
$\wdt{n}(\epsilon)$ for chaotic systems \cite{French:78} we obtain an
analytical expression for $\avg{P^{\delta}_k}$ in terms of
$\avg{P^n(k/N)}$; the final result is
\begin{equation}
\begin{split}
\avg{P^{\delta}_k} &= \avg{P^n_k}-\dfrac{1}{12},\;\;k= 1,2,\cdots,N-1,
\;\;N \gg 1.
\end{split}
\label{pdk_pnk}
\end{equation}
Eq. (\ref{pdk_pnk}), together with (\ref{pnk_pn}) and (\ref{GOE}) or
(\ref{GUE}), gives explicit expressions of $\avg{P^{\delta}_k}$ for GOE
and GUE. For generic chaotic systems we can apply GOE or GUE results
depending on their symmetries.

{\em Integrable systems.--} A similar calculation can be performed for
integrable systems. In this case $K(\tau)=1$ \cite{Stockmann:99}, and
instead of (\ref{GUE}) or (\ref{GOE}) we have $P^n(\tau)=\dfrac{1}{4
\pi^2 \tau^2}$ \cite{Robnik:03} for all $\tau$.  Using (\ref{pnk_pn}),
obtained with the same assumptions (a) and (b), and the fact that the
variances of $\delta_q$ and $\wdt{n}(\epsilon)$ are equal for
integrable systems, we get
\begin{equation}
\avg{P^{\delta}_k} = \dfrac{1}{4 \sin^2 \left(\dfrac{\pi k}{N} \right)},
\;\;N\gg 1.
\label{pdk_int}
\end{equation}

In order to check the assumptions (a) and (b) introduced above, we may
now give an alternative proof of eq. (\ref{pdk_int}) without using
these assumptions. For integrable systems, the spacings sequence $s_i$
can be considered as a sequence of independent random variables with
Poisson distribution \cite{Berry:77}. Therefore, the $\delta_q$
statistic is a sum of such random variables $\delta_q = \sum_{i=1}^q
w_i$, with $w_i = s_i - <s>$. Additionally, in this case we have 
$\avg{\wdh{w}_k}=0$ and $\avg{\left|\wdh{w}_k\right|^2}=1$, where
$\wdh{w}_k$ is the Fourier transform of $w_i$. Then, we can write
\begin{equation}
\delta_q = \dfrac{1}{\sqrt{N}}\sum_{k=0}^{N-1} \wdh{w}_k \sum_{m=1}^{q}
e^{-\dfrac{2 \pi i k m}{N}}.
\end{equation} 
This series can be added up considering that $\avg{\wdh{w}_k}=0$; we obtain
for $N\gg 1$
\begin{equation}
\delta_q = \dfrac{1}{\sqrt{N}}\sum_{k=0}^{N-1} \dfrac{i \wdh{w}_k e^{\dfrac{-i \pi k}{N}}}
{2\sin \left( \dfrac{\pi k}{N} \right)} e^{-\dfrac{2 \pi i k q}{N}},
\end{equation}
and therefore
\begin{equation}
\wdh{\delta}_k = \dfrac{i \wdh{w}_k e^{\dfrac{-i \pi k}{N}}}{2
\sin \left( \dfrac{\pi k}{N} \right)},\;\; N\gg 1.
\end{equation}
Consequently, taking into account that for a Poissonian sequence
$\avg{\left|\wdh{w}_k\right|^2}$=1, we can write
\begin{equation}
\avg{P^{\delta}_k} = \avg{\left|\wdh{\delta}_k\right|^2} =
\dfrac{1}{4 \sin^2 \left(\dfrac{\pi k}{N} \right)},\;\; N\gg1.
\label{pdk_int_2}
\end{equation}
The coincidence of eqs. (\ref{pdk_int}) and (\ref{pdk_int_2}) shows
that the assumptions (a) and (b) introduced to derive (\ref{pdk_pnk})
and (\ref{pdk_int}) are sound approximations.

{\em $1/f$ and $1/f^2$ noises.--} Note that eqs. (\ref{pdk_pnk}) and
(\ref{pdk_int}) are generic results, valid for every chaotic and
integrable quantum system, respectively. Nevertheless, there can be
some deviations at the lower frequencies due to the behavior of the
shortest periodic orbits. When $k\ll N$ the first term of
eq. (\ref{pnk_pn}) becomes dominant and using eq. (\ref{pn_smallt}) we
can write, for chaotic systems,
\begin{equation}
\avg{P^{\delta}_k} = \dfrac{N}{2 \beta \pi^2 k},\;\; k \ll N,\;\;N\gg1.
\label{1/f}
\end{equation}
Similarly, for integrable systems, eq. (\ref{pdk_int}) becomes
\begin{equation}
\avg{P^{\delta}_k} = \dfrac{N^2}{4 \pi^2 k^2},\;\; k \ll N,\;\;N\gg 1.
\label{1/f2}
\end{equation}
These expressions show that for small frequencies, the excitation
energy fluctuations exhibit $1/f$ noise in chaotic systems and $1/f^2$
noise in integrable systems. As we shall see below, these power laws
are also approximately valid through almost the whole frequency
domain, due to partial cancellation of higher order terms in
eqs. (\ref{pdk_pnk}) and (\ref{pdk_int}). Only near $k=N/2$, the so
called Nyquist frequency \cite{dft}, the effect of these terms becomes
appreciable.

\begin{figure}[h]
\begin{center}
\leavevmode
\psfig{file=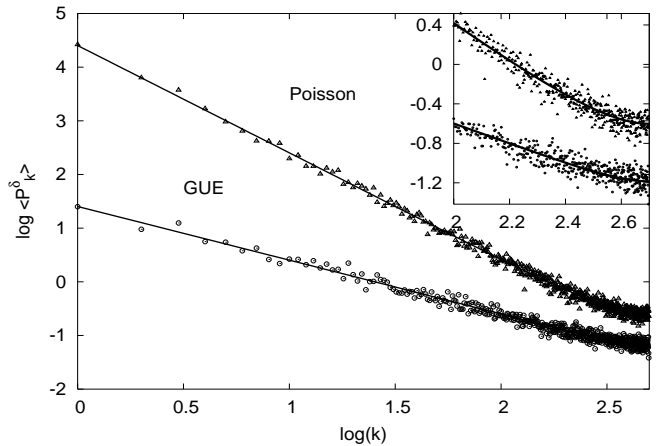,height=9cm,width=6cm,angle=-90}
\caption[]{Theoretical power spectrum of the $\delta_q$ function for
GUE and integrable systems (solid lines), compared to the results
calculated numerically using 30 GUE matrices of dimension $N=10^3$
(circles) and 30 Poisson level sequences of length $N=10^3$
(triangles).}
\label{fig:ensembles}
\end{center}
\end{figure}

To test all these theoretical expressions we have compared their
predictions to numerical results obtained for different ensembles and
systems.  Fig. \ref{fig:ensembles} displays the theoretical values of
$\avg{P^{\delta}_k}$ for GUE and integrable systems, as given by
(\ref{pdk_pnk}) and (\ref{pdk_int}), together with the numerical
results for GUE matrices and Poisson level sequences published in
\cite{Relano:02}. In order to enlarge the high frequency region, where
the numerical results show small deviations from the $1/f^{\alpha}$
power law behavior, an upper right panel is added to the figure.  The
agreement between the theoretical and numerical results is excellent
at all frequencies (note that there are no free parameters in the
analytical result). The theoretical curve describes perfectly the
power laws, characteristic of small and intermediate frequencies, as
well as the deviations observed at the highest frequencies.

\begin{figure}[h]
\begin{center}
\leavevmode
\psfig{file=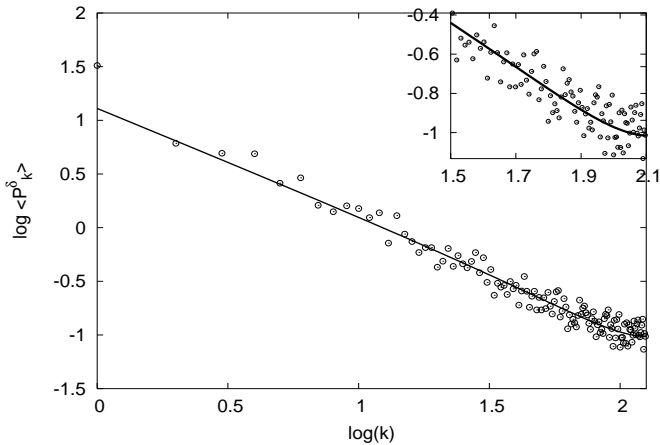,height=9cm,width=6cm,angle=-90}
\caption[]{Numerical average power spectrum of the $\delta_q$ function
for $^{34}$Na, calculated using 25 sets of 256 consecutive levels from
the high level density region, compared to the parameter free
theoretical values (solid line) for GOE.}
\label{fig:nucleo}
\end{center}
\end{figure}

\begin{figure}[h]
\begin{center}
\leavevmode
\psfig{file=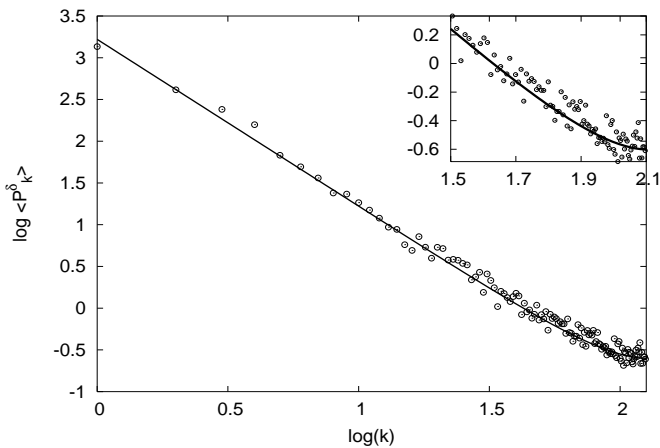,height=9cm,width=6cm,angle=-90}
\caption[]{Numerical average power spectrum of $\delta_q$ for a
rectangular billiard, calculated using 25 sets of 256 consecutive
levels, compared to the parameter free theoretical values (solid line)
for integrable systems.}
\label{fig:billar}
\end{center}
\end{figure}

We have also compared our predictions to the power spectra of
$\delta_q$ for two physical systems: an atomic nucleus (chaotic) and a
rectangular billiard (integrable).

In the first case we have performed a shell model calculation for
$^{34}$Na using an adequate realistic interaction. The Hamiltonian
matrices for different angular momenta, parity and isospin were fully
diagonalized. Then, 25 sets of 256 consecutive high energy levels of
the same $J^{\pi} T$ were selected, and the average power spectrum of
the $\delta_q$ function was calculated. Fig. \ref{fig:nucleo} shows
the result of this calculation together with the theoretical values of
eq. (\ref{pdk_pnk}). An excellent agreement between the theoretical and
numerical results is obtained through the whole frequency interval.

As an example of integrable system we have chosen a rectangular
billiard with sides of length $a=\sqrt{\lambda}$ and
$b=1/\sqrt{\lambda}$, with $\lambda=(\sqrt{5}+1)/2$; this geometry
gives rise to an irrational ratio $a/b=\lambda$, and thus there are no
degeneracies in the spectrum. We have calculated the spectrum and
selected 25 sets of 256 consecutive very high energy levels in order
to avoid, as far as possible, the influence of short periodic
orbits. Fig. \ref{fig:billar} shows the results for the average power
spectrum of $\delta_q$ and the theoretical values given by eq.
(\ref{pdk_int}). As in previous cases, we can see that the agreement
between the theoretical and numerical results is very good in the
whole frequency domain.

In summary, we have derived theoretical expressions for the power
spectrum of the $\delta_q$ function both for regular and chaotic
quantum systems. These expressions are universal and do not contain
any free parameter. We have compared our theoretical predictions with
numerical results for RME, a rectangular billiard and an atomic
nucleus, obtaining an excellent agreement for all these systems.  The
theory reproduces the power laws of type $1/f$ for chaotic systems,
and $1/f^2$ for regular ones, observed in the power spectrum of the
excitation energy fluctuations up to frequencies very close to the
Nyquist limit.

We are particularly indebted to P. Leboeuf, O. Bohigas and M. Robnik
for enlightening discussions.  This work is supported in part by
Spanish Government grants BFM2003-04147-C02 and FTN2003-08337-C04-04.

\end{document}